\documentclass[twocolumn,trackchanges]{aastex62} 

\usepackage{color}
\graphicspath{{./}{figures/}}

\received{???}
\revised{???}
\accepted{???}
\shortauthors{Wang et al.}

\begin{document}

\title{Age Determination of LAMOST Red Giant Branch stars based on the Gradient Boosting Decision Tree method}
\author[0000-0001-8459-1036]{Hai-Feng Wang}
\affil{Dipartimento di Fisica e Astronomia ``Galileo Galilei", Universit\'a degli Studi di Padova, Vicolo Osservatorio 3, I-35122, Padova, Italy}
\author[0000-0002-0155-9434]{Giovanni Carraro }
\affil{Dipartimento di Fisica e Astronomia ``Galileo Galilei", Universit\'a degli Studi di Padova, Vicolo Osservatorio 3, I-35122, Padova, Italy}
\author{Xin Li}
\affil{Department of Astronomy, China West Normal University, Nanchong, 637002, P.\,R.\,China}
\author{Qi-Da Li}
\affil{Yunnan Observatories, Chinese Academy of Sciences, Kunming 650216, P.\,R.\,China}
\author[0000-0001-5831-1889]{Lorenzo Spina}
\affil{INAF-Padova Observatory, Vicolo dell'Osservatorio 5, 35122 Padova, Italy}
\author{Li Chen}
\affil{Shanghai Astronomical Observatory, Chinese Academy of Sciences, 80 Nandan Road, Shanghai 200030, P.\,R.\,China}
\author{Guan-Yu Wang}
\affil{Department of Astronomy, China West Normal University, Nanchong, 637002, P.\,R.\,China}
\author{Li-Cai Deng}
\affil{National Astronomical Observatories, Chinese Academy of Sciences, Beijing 100101, P.\,R.\,China}

\begin{abstract}
In this study we estimate the stellar ages of LAMOST DR8 Red Giant Branch (RGB) stars based on the Gradient Boosting Decision Tree algorithm (GBDT). We used 2,643 RGB stars extracted from the  APOKASC-2 astero-seismological catalog as  training data-set. After selecting the parameterses ([$\alpha$/Fe], [C/Fe], T$_{eff}$, [N/Fe], [C/H], log g) highly correlated with age using GBDT, we apply the same GBDT method to the new catalog of more than 590,000 stars classified as RGB stars. The test data-set shows that the median relative error is around 11.6$\%$ for the method. We also compare the predicted ages of RGB stars with other studies (e.g., based on APOGEE), and find systematic differences. The final uncertainty is about 15 to 30$\%$ compared to open clusters' ages. Then we present the spatial distribution of the RGB sample having an age determination, which could recreate the expected result, and discuss systematic biases. All these diagnostics show that one can apply the GBDT method to other stellar samples to estimate atmospheric parameters and age.  

\end{abstract}

\keywords{Milky Way disk (1050); Stellar ages (1581); Red Giant Branch (1368); Catalogs (205)}

\section{Introduction} 

The Milky Way (MW) is a unique laboratory for Galactic archaeology studies because  a large number of individual stars can be precisely resolved. One can use multidimensional information to study the properties of stellar populations and the history of our Galaxy. The abundance of chemical elements on the surface of stars can be used to provide the fossil evidence of the Galactic environment at the time of their birth. In turn, stellar ages are helpful to trace back the history of the MW. Therefore, it is very essential to estimate reliable stellar ages for widely distributed stars across the MW \citep{2016ApJ...831..139M, 2019NatAs...3..932G, 2022arXiv220402989C, 2022Natur.603..599X, 2023arXiv230408276A, 2024MNRAS.528L.122C}.

Red Giant Branch (RGB) stars are at one stage occurring after the main sequence phase, which are the portion of the giant branch before helium ignition occurs in the course of stellar evolution. Red Clump Giant (RCG) stars typically form a distinct horizontal branch on the HR diagram which can be mixed with the RGB. Most of these stars are burning helium in their core (Primary RC or RCG). In terms of stellar populations, Red Giant Branch(RGB) stars are particularly beneficial for Galactic archaeology \citep[e.g.,][]{2015ApJ...809L...3S,2023arXiv230408276A}. They are numerous and cover a wide range of stellar ages \citep{2012ASSP...26.....M}. Because of their high luminosity, RGB stars can be observed at larger distances, thus probing all the way to the outskirts of our Galaxy. Moreover, their solar-like oscillations may be observed at longer frequencies than those required to observe similar oscillations in dwarf and sub-giant stars \citep{2023arXiv230410654S}.

Accurate stellar age estimation for large samples is essential for a comprehensive understanding of structure, kinematics, and dynamics of different stellar populations in the Milky Way \citep{2019MNRAS.489..176M, 2022MNRAS.512.4697L, 2019ApJ...883..177N, 2018A&A...612L...8L, 2018MNRAS.475.5487S,2021ApJ...909..115C,wang2018a,wang2018b,wang2019,wang2020a,wang2020b,wang2020c,wang2022,wang2023a,wang2023b,Li2023}. However it is difficult to estimate stellar ages directly from the observations, and one has to rely on empirical formulae \citep[e.g.,][]{2010ARA&A..48..581S}. To circumvent all this, often the abundance ratio [$\alpha$/Fe] is used as age proxy \citep{2013A&ARv..21...61R, 2016ApJ...823...30B, 2023A&A...669A.104K}.

Stellar evolution models have often been employed to estimate stellar ages by comparing stellar parameters inferred from observations with model predictions. A practical approach to age estimation of individual field stars is to use spectroscopic and/or photometric methods \citep{2023A&A...674A..27A, 2008A&A...486..951G, 2005ESASP.576..565B, 2017NatAs...1E.184S, 2008MNRAS.389...75B, 1999PASP..111...63F, 2023arXiv230303420Z}, to derive stellar photospheric parameters and compare the inferred star positions in the Hertzsprung–Russell diagram with theoretical tracks and/or isochrones \citep{2016A&A...585A..42S, 2017A&A...604A.108M, 2018MNRAS.477.5279M, 2018MNRAS.481.4093S, 2020A&A...642A..88L}. However, it is not easy to apply this method to RGB giants because the isochrones of giant stars of different ages suffer from severe mixing effects on the Hertzsprung–Russell diagram and are not precise enough \citep{2010ARA&A..48..581S,2015ASSP...39..167N}. Another effective method (but also dependent on models) is provided by asteroseismology \citep{2013ARA&A..51..353C}, which can be used to estimate the age of field stars including red giants. However, it has not been applied yet to large enough samples.

With the advent of surveys such as APOGEE \citep{2017AJ....154...94M}, many measurements of stellar parameters become available which can be used to estimate ages. The stellar surface properties log g and  T$_{eff}$ can be used to characterize stellar spectroscopic type and luminosity class. Then, in combination with theoretical stellar evolution models, one may obtain information on stellar ages \citep{2012MNRAS.427..127B, 2016JPhCS.703a2002S, 2016ApJ...823..102C}. However, stellar evolution induces changes in the surface chemical abundances through physical processes such as ``dredge-up", where CNO products (nitrogen-rich and carbon-poor) are brought up into the stellar atmosphere. These CNO processes  depend on mass, and therefore abundance ratios like [C/N] can provide deep insights into stellar evolution and can be used to estimate age \citep{2015A&A...583A..87S}. Thus, information about stellar evolution (i.e., age) is also correlated with [Fe/H], [$\alpha$/Fe], and other chemical abundances \citep{2015A&A...579A..52N, 2016A&A...593A..65N, 2017A&A...608A.112N, 2016A&A...590A..32T, 2017MNRAS.465L.109F, 2018MNRAS.474.2580S, 2019A&A...629A..62C, 2020A&A...633L...9J, 2021A&A...652A..25C, 2022A&A...660A.135V}. Finally, the measured stellar abundances depend on the material from which the star formed \citep{2016A&A...593A.125S, 2019A&A...624A..78D, 2019ApJ...883..177N}.

\citet{2016ApJ...823..114N} and \citet{2016yCat..74563655M} determined stellar masses from spectroscopy, which greatly extends the range of giant stars with age estimates. They showed that the masses (and implied age \citep{2015MNRAS.453.1855M, 2017A&A...601A..27L, 2020ApJS..249...29H}) of post dredge-up giants can be measured from high-resolution infrared spectra (APOGEE \citep{2019PASP..131e5001W}, R $\approx$ 22,500), and the correlation of  masses and ages with T$_{eff}$, log g, [M/H], [C/M] and [N/M]. This study has increased the number of giant stars of known age to 70,000.

\citet{2015ApJ...808...16N} described a data-driven stellar modeling method called ``Cannon" that allows for spectral mass measurements and \citet{2017ApJ...841...40H} used Cannon to transfer labels from a high-resolution, high-S/N survey (APOGEE) to a low-resolution, medium-S/N survey (LAMOST). The author showed that using Cannon, the fundamental parameters (T$_{eff}$, log g, [Fe/H] and [$\alpha$/M]) consistent with the APOGEE values can be determined directly from the LAMOST spectra. They applied the model to 450,000 giants from LAMOST DR2 that were not observed by APOGEE, and this dramatically increased the number and sky coverage of stars with mass and age estimates. \citet{2018ApJ...858L...7T} has also estimated us 175,202 red clump stars in LAMOST with 3\% contamination, and also includes two asteroseismology parameters $\Delta$P and $\Delta\nu$.

However, until the arrival of the PLATO mission \citep{2014ExA....38..249R, 2017AN....338..644M}, red giants ages from the common constraints of asteroseismology and spectroscopy were only applicable to specific samples in certain fields, such as Kepler \citep[e.g.,][]{2014ApJS..215...19P, 2018ApJS..239...32P, 2018MNRAS.475.3633W, 2021yCat..36450085M}, CoRoT \citep{2016AN....337..970V, 2017A&A...597A..30A}, K2 \citep[e.g.,][]{2019MNRAS.490.4465R, 2022ApJ...926..191Z}, or TESS continuum observing regions \citep{2018MNRAS.473.2004S,2020ApJ...889L..34S,2021MNRAS.502.1947M, 2023MNRAS.520.1913W}. Thus, large-scale spectroscopic surveys like APOGEE \citep{2017AJ....154...94M}, GALAH \citep{2015MNRAS.449.2604D} or LAMOST \citep{2012RAA....12.1197C}  have been dedicated to providing empirical spectroscopic-based age estimates for Galactic archaeological studies, and asteroseismic data are simply used as benchmarks \citep[e.g.,][]{2016yCat..74563655M, 2019MNRAS.483.3255L, 2022MNRAS.512.1710H}. Such large samples of stellar spectroscopic ages provide great scientific value for Galactic studies, even if their accuracy is below the ideal requirement \citep{2019ApJS..245...34X}(which is inherited from both the Payne \citep{2019ApJ...879...69T}). In this work, we use the stellar ages analyzed by \citet{2018ApJS..239...32P} based on APOGEE spectral parameters and Kepler asteroseismic data as the training sample.

Asteroseismology is considered to be one of the most accurate ways to obtain stellar ages so far. Currently, spectral ages obtained from the asteroseismology dataset have a number of limitations. \citet {2023MNRAS.522.4577L} proposed to address these issues by applying a variational encoder-decoder on cross-domain astronomical data. The model was trained on stellar pairs from both APOGEE and Kepler observations, these APOGEE spectra can then be trained to predict ages with about 1000 precise asteroseismic stellar ages. The model produces more accurate spectral ages for APOGEE DR17 ($\sim$ 22 $\%$ overall and $\sim$ 11 $\%$ for red clump stars) compared to previous data-driven spectral ages. However, due to the lack of asteroseismic parameters, we have chosen to use machine learning algorithms in our work to determine stellar ages using correlations between stellar feature parameters and ages.

As for machine learning, one can rely on a training set having high quality stellar parameters and then apply the tool to predict a large sample of stellar ages. So far, many astronomical data measurements are  based on  machine learning algorithms \citep{2021MNRAS.503.2814C, 2019ApJ...879...69T, 2019ApJ...878...21T, 2022MNRAS.517.5325H}. Using the third data release of the GALAH survey, \citet{2021MNRAS.506..150B} recently have demonstrated that by using supervised machine learning regression, specifically the popular eXtreme Gradient Boosting method (XGBoost; \citet{2016arXiv160302754C}), it is possible to infer the ``spectrum" (or ``chemistry") of main-sequence-turn-off stellar ages with an accuracy of 1$-$2 Gyr. \citet{2022MNRAS.512.1710H} performed a similar exercise using the same technique, but employing red clump stars observed by the LAMOST survey (which has the advantage of providing C and N abundances), trained on Kepler's asteroseismic ages. These authors derived a statistical uncertainty of 31 $\%$. Similar accuracy was obtained earlier by \citet{2019MNRAS.489..176M} using a Bayesian convolution neural network and APOGEE DR14 data trained on APOGEE-Kepler data \citep{2018ApJS..239...32P}. In order to reveal the formation pathways of the inner and outer disk, \citet{2021MNRAS.503.2814C} developed a Bayesian Machine Learning framework for the age determination, which also is conditioning on the APOGEE and Kepler asteroseismic age data.

\citet{2022ApJS..262...20L} used the random forest (RF) methods and the convex-hull algorithms to predict RGB (Red Giant Branch) and RCG ((Red Clump Giants)) masses based on the scikit-learn \citet{Pedregosa2011}, and hence RCG ages. The median relative error is 13 $\%$ for large sample of K giants masses, and 9 $\%$ and 18 $\%$ for RC stellar masses and ages, respectively. When compared with the open cluster age, the uncertainty is about 10 $\%$. And for the first time, to the best of our knowledge, they quantitatively compare different machine learning methods and find that nonlinear models are generally better than linear models.
Therefore the aim of our present study is to develop new algorithms for determining the RGB ages of a large sample of stars, while attempting at improving the accuracy of age predictions as much as possible. 
In this paper we use the stellar ages in APOKASC-2 as a training sample and estimate the ages of more than 590,000 stars classified as RGB stars from LAMOST DR8 using the GBDT.

The paper is structured as follows: in Section 2, we present the data used in this paper. In Section 3, the specific application of the GBDT algorithm to our work is presented. Section 4 shows our age prediction results and compares the results of this work with the literature, discussing the advantages and disadvantages of different methods. We will discuss the applications of the sample and the possibility of predicting age with other feature parameters in Section 5. Section 6 summarizes our conclusions. 

\section{Data} 

The results of a machine learning regressor are much dependent on the quality of the training dataset. Therefore, it is better to use a small but statistically significant and high quality dataset during the training stage \citep{2017A&A...597A..30A}. \citet{2018ApJS..239...32P} provided stellar parameters for 6,676 evolved stars in APOKASC-2, which includes stellar ages. These ages come from their model using mass, radius, [Fe/H], and [$\alpha$/Fe]. They adopted a new combination of five methods of asteroseismic measurements to calibrate the inferred stellar parameters. The precision is good enough for us to predict ages and have already been widely used in the community. In this paper, we use stars marked as RGB in APOKASC-2 to train the age prediction model.

We selected 3,777 stars marked as RGB in the APOKASC-2, and then we cross-match with LAMOST DR8 to get the sample of 2,643 stars, we exclude some stars with large errors using the GBDT prediction first, so only high-precision sources being retained, we finally get 2,087 stars for which 1,047 for the training and 1,040 for testing during the machine learning process.

\begin{figure}
  \centering
  \includegraphics[width=0.45\textwidth]{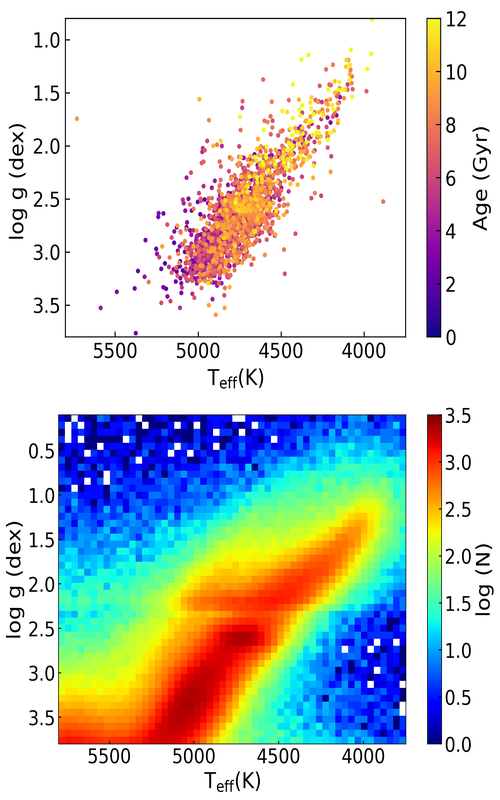}
  \caption{Teff-log g diagram of the sample used in this paper. The upper panel is the Kiel diagram of the RGB stars selected from the APOKASC-2 catalog as the final training small sample, and the colours indicate the stellar ages. The lower panel shows the Kiel diagram of the large sample to be estimated using the GBDT method. The sample of RGB is identified by \citet{2023A&A...675A..26W} from LAMOST DR8 since our target is to test the method during this work, with a purity and completeness exceeding 95$\%$.}
  \label{RGB_Sample_distribution}
\end{figure}

\begin{figure}
  \centering
  \includegraphics[width=0.45\textwidth]{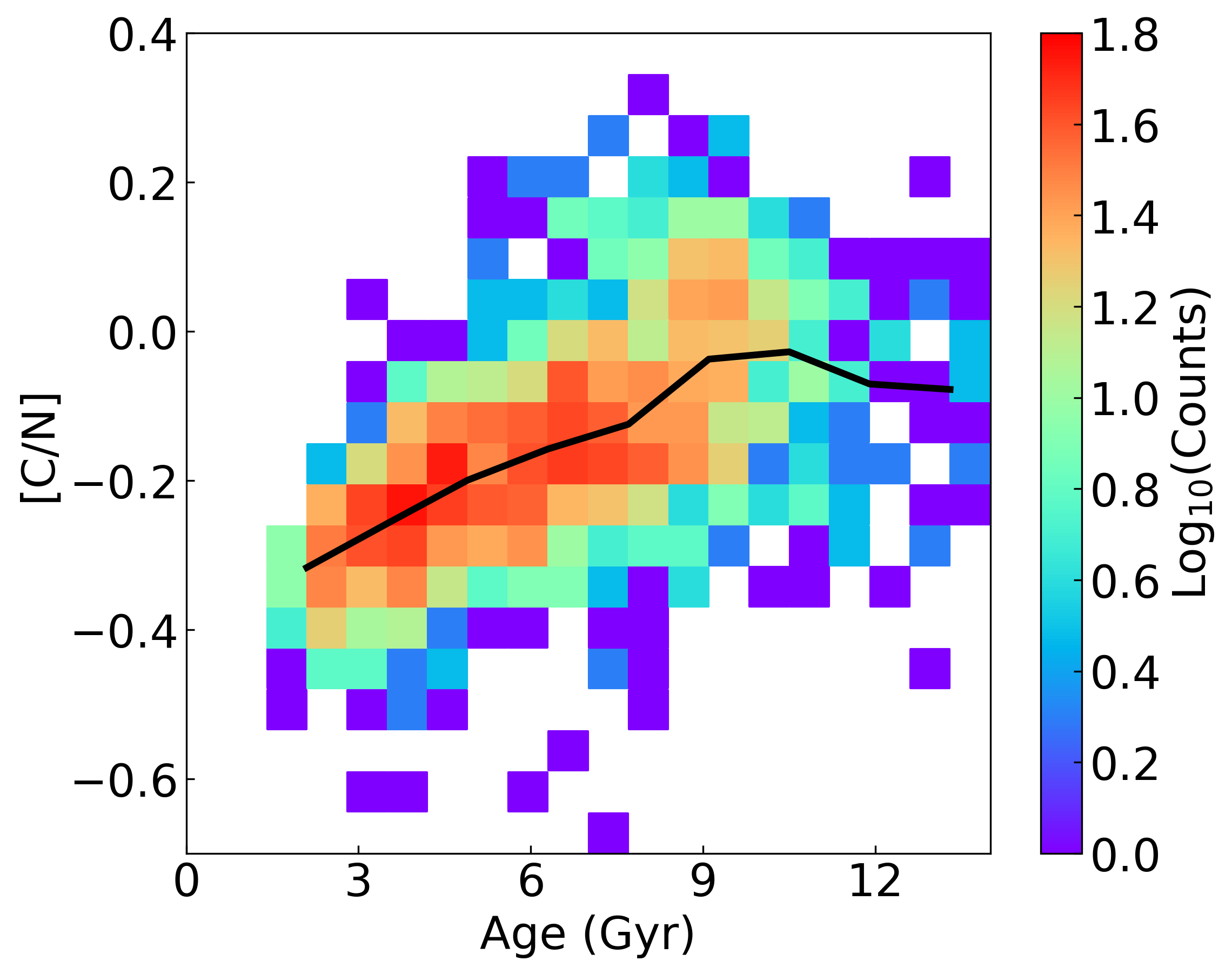}
  \caption{[C/N]-age diagram of the training sample used in this paper and the colour indicates the stellar density in the log scale, it shows that both parameters are correlated with each other.}
  \label{Train_Age_C_N_Counts}
\end{figure}

As one can see in Figure~\ref{RGB_Sample_distribution}, the top panel shows the Kiel plot of our RGB samples for the final high-resolution asteroseismology dataset, with the colours indicating the high accurate age. The bottom panel contains the RGB targets for age determination from LAMOST DR8, where it is clear that the surface gravity and effective temperature are consistent with the K type RGB stars. As a complement, we also show the age and carbon to nitrogen ratio (C/N) pattern for the same sample (APOKASC-2) in Figure~\ref{Train_Age_C_N_Counts}, which shows good correlation up to 10 Gyr. However, the statistical data quality deteriorates beyond this age threshold.

\citet{2022ApJS..259...51W} provided a value-added catalog for LAMOST DR8 \footnote{In principle we can condition on the criteria of T$_{eff}$ and log g in \citet{2014ApJ...790..110L} to select the DR11 K giants, and choose or not to separate into the RGB and RCG, then determine the basic abundance and finally get the age, which has been done. But as known, it should almost be another a piece of work. During this work we just purely want to explore and test the method for the age determination, so we choose to adopt the publicly available dataset for RGB.}, which includes 7.1 million stellar parameters estimated from low-resolution spectra. The catalog provides the values of stellar atmospheric parameters (effective temperature T$_{eff}$, surface gravity log g, metallicity [Fe/H]/[M/H]), the $\alpha$-element-to-metallicity ratio [$\alpha$/M], carbon-to-iron abundance ratios [C/Fe] and [N/Fe], and other stellar parameters.
In this study we adopt the chemical abundances of \citet{2022ApJS..259...51W} to test our new method. Note that there might have chemical outliers in that work, but it is not the topic of this paper. We are based on the APOKASC-2 to train the model for age prediction.

\citet{2023A&A...675A..26W} selected 696,680 red giant branch (RGB) stars, 180,436 primary red clump (RC) stars, and 120,907 secondary red clump (SRC) stars from LAMOST DR8 based on large frequency spacing ($\Delta$$\nu$) and period spacing ($\Delta$P). The purity and completeness of the RGB and RC samples are both over 95$\%$ and 90$\%$, respectively. We used the RGB provided by \citet{2023A&A...675A..26W}, but the age prediction-related parameters needed in this paper are almost all null or spurious in some of the stars. The cleaning of the sample produced 596,116. The Kiel diagrams of all the remaining stars are also shown in the lower panel of Figure~\ref{RGB_Sample_distribution}, with the colours indicating the density of the stars in a log scale. We apply the GBDT Machine Learning trained age prediction model to this large catalog to provide reliable ages.

\section{Method} 

\subsection{Gradient Boosting Decision Tree}

In this paper, we adopt the Gradient Boosting Decision Tree algorithm \citep{friedman1999} based on the scikit-learn python package \citep{Pedregosa2011}, which is an additive model based on the idea of boosting integrated learning. It is an iterative decision tree algorithm also called MART (Multiple Additive Regression Tree). It works by constructing a set of weak learners (trees) and accumulating the results of multiple decision trees as the final prediction output. Weak learners often refer to learners that generalize slightly better than randomly generated results (e.g., classifiers with slightly more than 50\% accuracy on binary classification problems). The algorithm effectively combines decision trees with integration ideas.

Decision tree is a basic classification and regression method. The model consists in a tree structure, which can be considered as a collection of the so called ``if-then" rules. At the same time, decision tree algorithms require less feature engineering than other algorithms, since it can handle data with missing fields well. Decision trees are able to combine multiple features automatically. And the multiple decision trees are integrated by gradient boosting, which can eventually solve the over-fitting problem well. The hyper-parameters can be found in the the publicly available package. 

For GBDT, we set the parameters n\_estimators = 100, loss = ls, learning\_rate = 0.1, subsample = 1, criterion = friedman\_mse, max\_depth = 3, min\_samples\_leaf = 1, min\_samples\_split = 2, max\_features = None. n\_estimators specifies the number of weak classifiers. Increasing this value can improve accuracy, but beyond a certain point, the improvement becomes limited, loss is used to specify the loss function, where 'ls' stands for least squares regression, learning\_rate is used to adjust the contribution of each tree, subsample is used to fit the number of samples for individual learners, criterion measures the regression effects criterion, 'friedman\_mse' indicates improved mean square error, max\_depth specifies the maximum depth of individual regression estimates, min\_samples\_leaf specifies the minimum number of samples required in a leaf node, min\_samples\_split specifies the minimum number of samples required for a split at each internal node (non-leaf node), max\_features limits the number of features to be considered when finding the optimal split. 

\subsection{Feature parameter selection}

\begin{figure}
  \centering
  \includegraphics[width=0.45\textwidth]{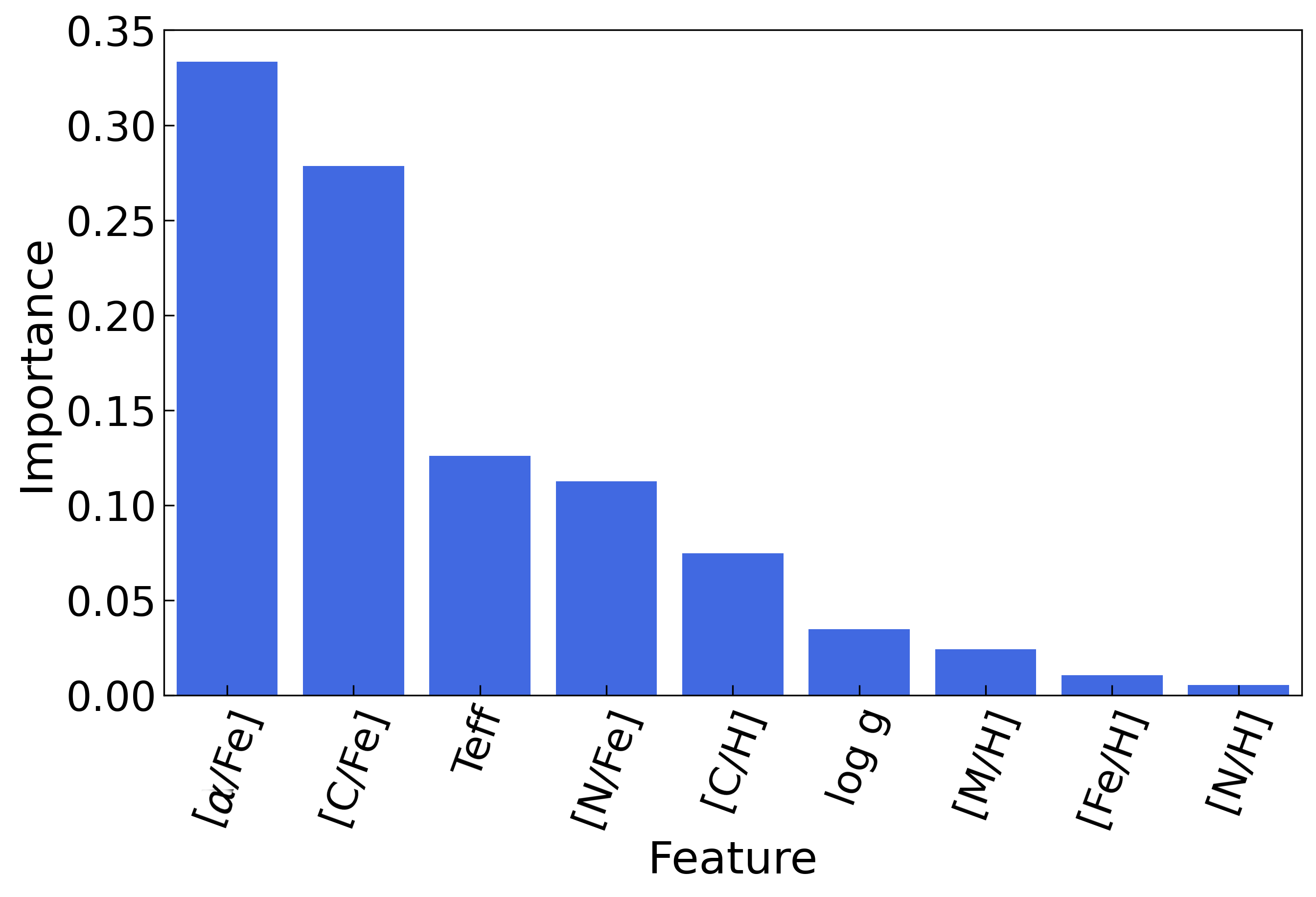}
  \caption{Stellar feature parameters extracted using the gradient boosting decision tree algorithm. The stellar features are sorted from left to right in order of importance, with the importance reflecting the contribution of that feature in the predicting model.}
  \label{Feature_importances}
\end{figure}

\begin{figure}
  \centering
  \includegraphics[width=0.45\textwidth]{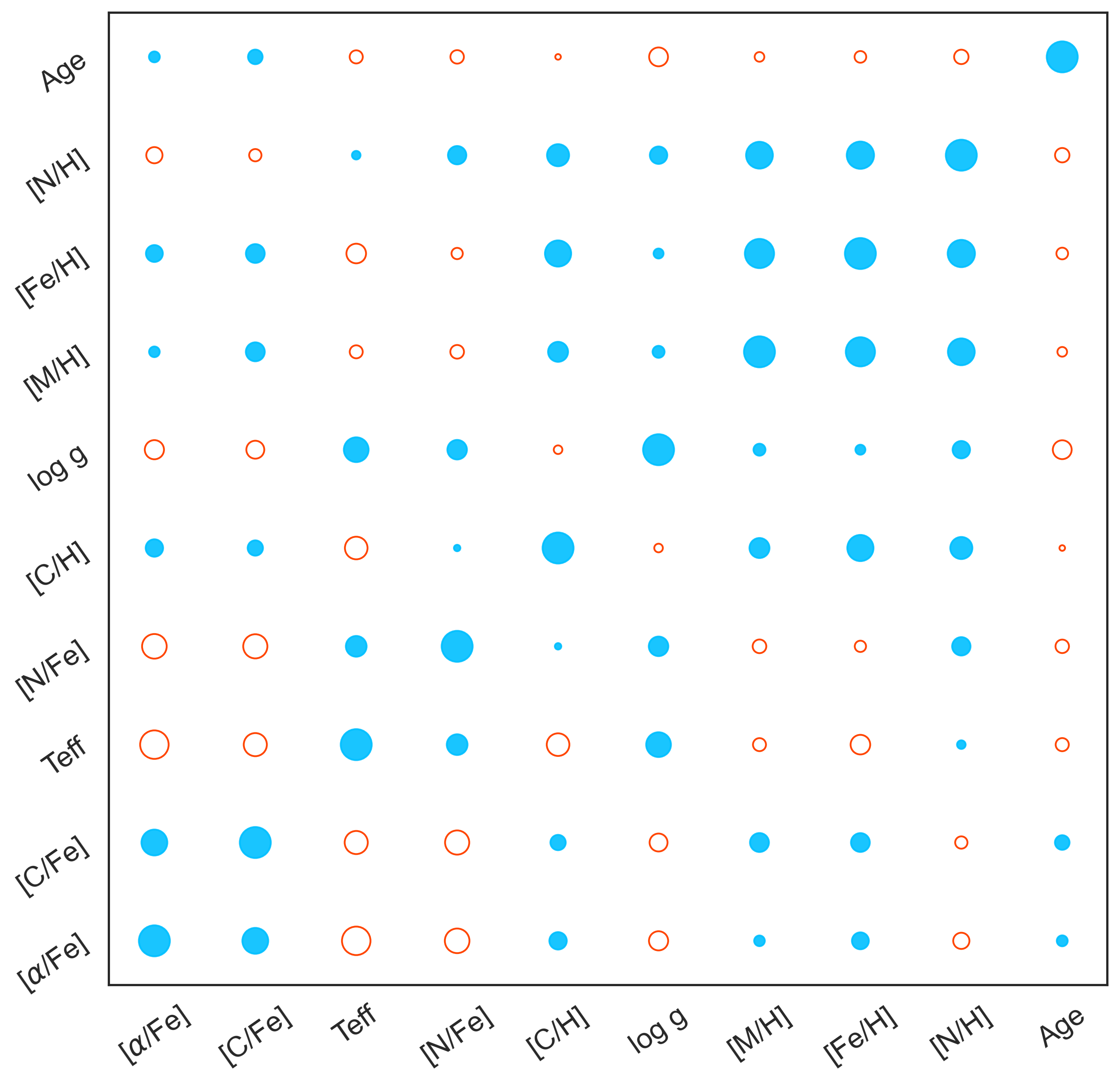}
  \caption{Correlation matrix for all parameters of the training dataset, with blue indicating positive correlation and orange indicating negative correlation. Note that it is for the training sample and we reasonably choose to condition on the importance analysis to estimate age as shown in Figure~\ref{Feature_importances}.}
  \label{correlation_matrix}
\end{figure}

We first extract the importance of the features in the training samples using GBDT, normalized them, and produced the sorting shown in Figure~\ref{Feature_importances}. The importance of the features reflects the contribution of that feature in the predicting model, from left to right, [$\alpha$/Fe], [C/Fe], T$_{eff}$, [N/Fe], [C/H], log g, [M/H], [Fe/H], [N/H]. However, these degrees of importance are relative, not absolute. All feature parameters also have some correlation with each other as shown in Figure~\ref{correlation_matrix}. Nonetheless,  we make use of Figure~\ref{Feature_importances} to estimate the age in this study. Therefore, it is not necessary to use all the feature parameters for model training, but rather select the first few highly correlated ones from them as a reasonable input.

\begin{figure}
  \centering
  \includegraphics[width=0.45\textwidth]{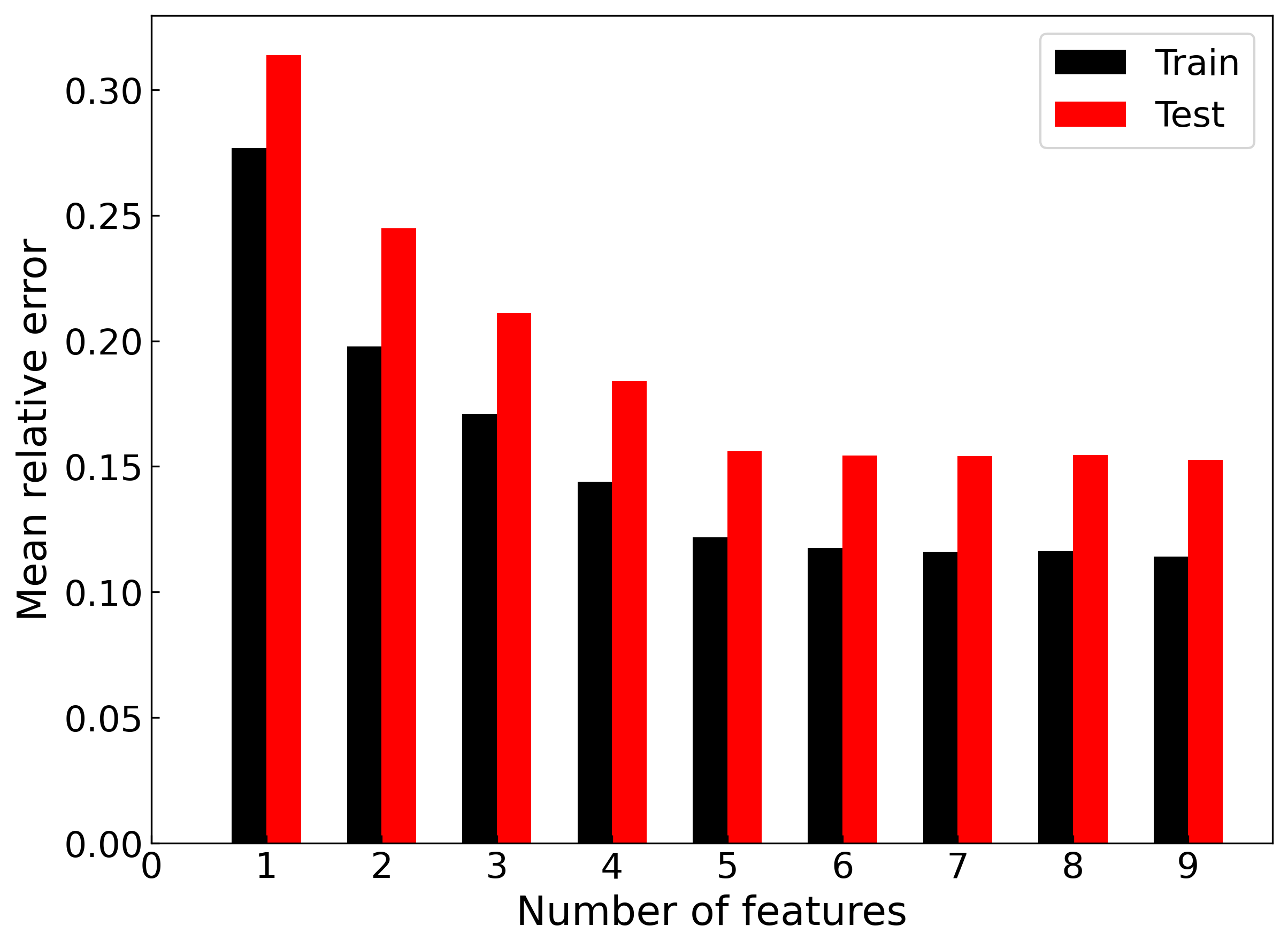}
  \caption{The average relative error between the training and test datasets as the number of training stellar features varies. The black histograms indicate the training dataset and the red histograms indicate the test dataset. The figure indicates the average relative error of the most stable value when there are 6 features.}
  \label{NOF_MRE_GBDT}
\end{figure}

It is found that there is a high correlation between all these stellar features by test (discussed in the appendix section), so it is feasible to select the top few features in the order of importance to predict the age. Therefore we will increase the number of features in the model for age prediction, and calculate the average relative error each time. The average relative error at different number of features is shown in Figure~\ref{NOF_MRE_GBDT}. It can be seen that the average relative error decreases as the number of stellar features increases. Therefore, we finally select the first six features ([$\alpha$/Fe], [C/Fe], T$_{eff}$, [N/Fe], [C/H], log g) for the training of the predicted age model.

\subsection{Performance on the test dataset}

\begin{figure}
  \centering
  \includegraphics[width=0.45\textwidth]{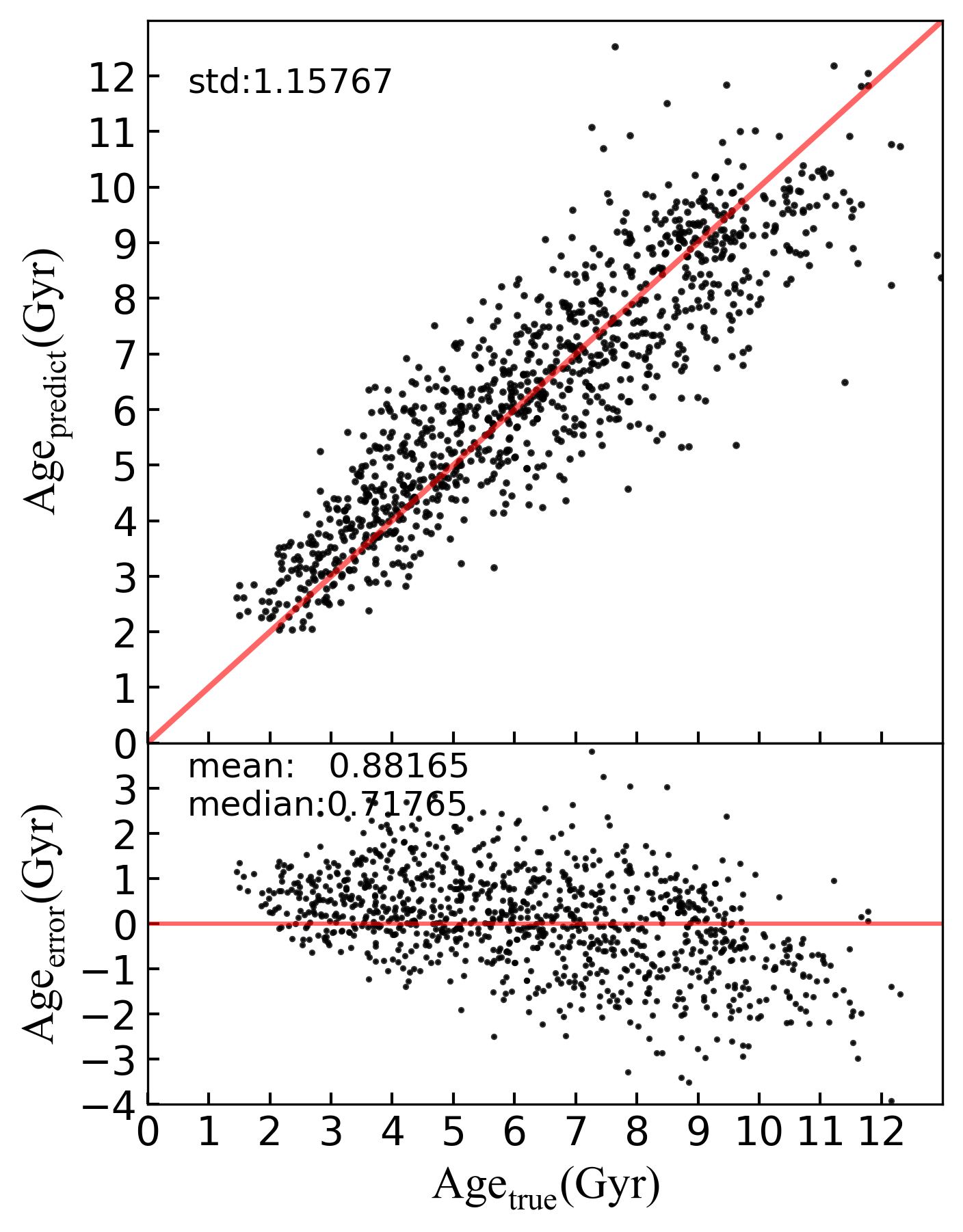}
  \caption{Errors analysis of the predicted ages for the test sample. The black dots are the ages of the test sample, x-axis is the true value and y-axis is the predicting value, and the red solid lines guide our eyes for performance of the method. The dispersion is calculated by subtracting the predicted age from the literature age in the test set and then taking the standard deviation. The absolute error is the predicted value minus the true value.}
  \label{True_Predict_Error_GBDT}
\end{figure}

The results of our model on the test dataset are shown in Figure~\ref{True_Predict_Error_GBDT}. The upper panel shows the predicted age compared to the APOKASC-2 test sample age and the dispersion, and the lower panel shows the absolute age error versus the true age and the mean and median of the absolute errors. We see that in the comparison of predicted age to original age, the scatter points and the red line exhibit a similar trend, indicating that the predicted age coincides with the true age. The points that do not fall on the solid red line are also evenly distributed on both sides of the line, and the dispersion of the predicted age is 1.16 Gyr. The absolute error has a similar distribution, with a mean and median of 0.88 Gyr and 0.72 Gyr. Note that in order to ensure that the range of parameter distributions in both the training dataset and the test dataset are uniformly covered with respect to the range of the total sample distribution, we adopt the following data partitioning method: we randomly extract the star from the training and test dataset, trying to keep both datasets almost equal.

\begin{figure}
  \centering
  \includegraphics[width=0.45\textwidth]{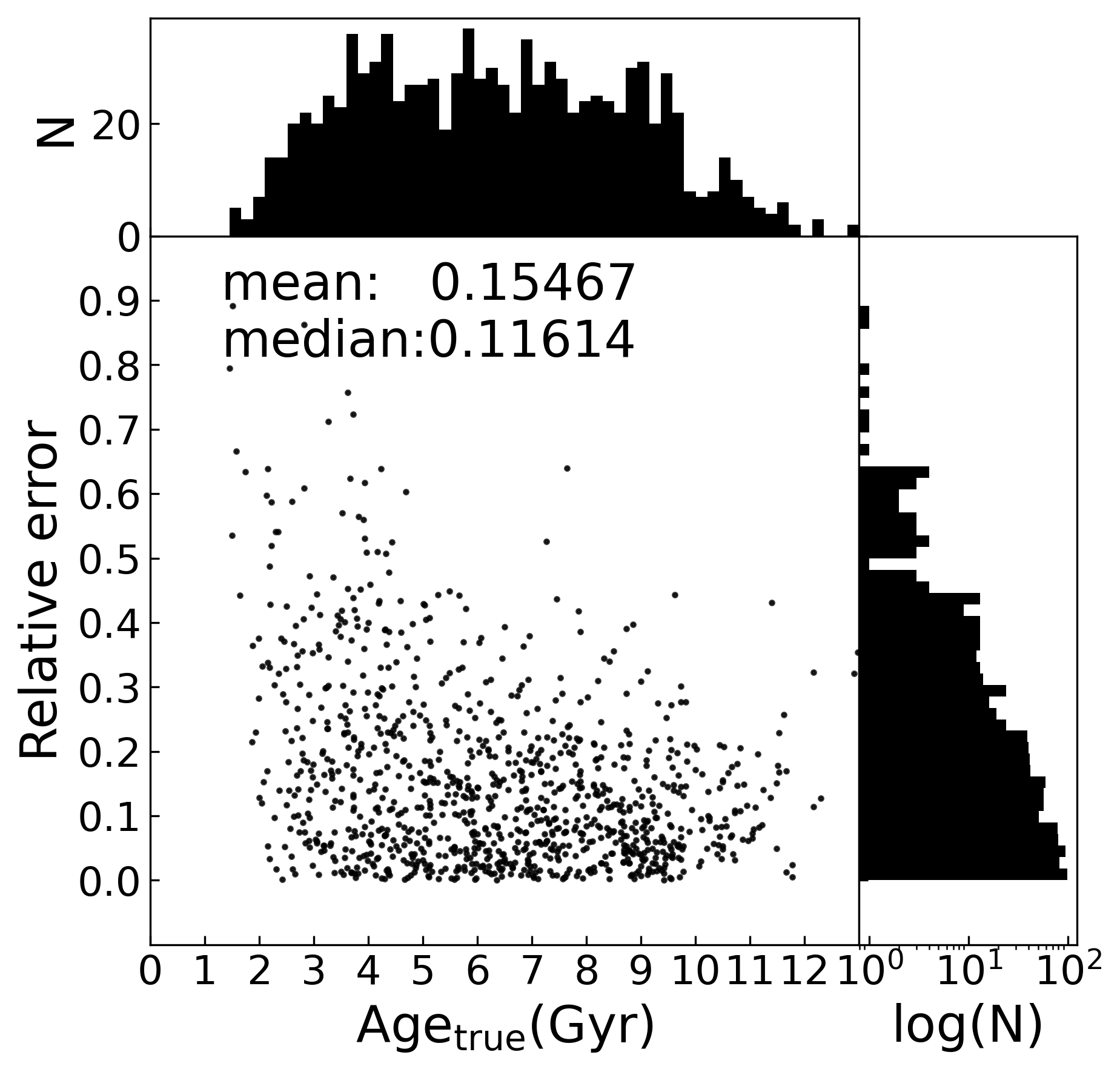}
  \caption{The mean and median relative error of predicted age for the test dataset. The black dots represent the relative error in age, the top histogram is the star counts, and the right panel shows the distribution of the relative age error in terms of values. The relative error is the absolute error divided by the true values. It is naturally found that the errors should be small since they are from the same catalog but divided as test and training.}
  \label{True_Relative_error_GBDT}
\end{figure}

In addition, in Figure~\ref{True_Relative_error_GBDT}, we show the mean\footnote{As the standard mathematical  definition.calculating the residuals between the predicted value and the “true value” of the original datasets, then using the residuals divided by the true value.} and median relative errors of the ages predicted by the model for the test dataset, as well as the distribution of the relative age errors with age. The relative error is the absolute error derived above divided by the true value. The average relative error of our predicted stellar ages is 15 $\%$, with a median relative error of 12 $\%$. Because machine learning will autonomously search for relevant patterns from the training sample, if we can have a much larger number of training samples with higher accuracy, then our prediction accuracy can be further improved. 

The top histogram of Figure~\ref{True_Relative_error_GBDT} shows that the ages of our training sample are almost homogeneously distributed, with a smaller number of older stars. We also note that in the lower panel of Figure~\ref{True_Predict_Error_GBDT}, the prediction errors for older stars are often below the red line, implying that our age predictions for older stars may be weak, in line with what can be found in other studies. The possible reasons are: the training dataset may have a limited number of samples for older stars, or the features of these samples may not be accurately or comprehensively represented, causing the model to fail to accurately learn and understand the features of older stars, resulting in underestimated age predictions. Insufficient feature selection is also a factor. The performance of GBDT model heavily relies on the selection and representation of features. If the selected features fail to capture the key characteristics of older stars, or the feature representation is not comprehensive enough, the model may struggle to accurately predict the age of older stars.

For the future we plan to use a new machine method to extract pure RGB stars and exploit a larger suite of chemical abundances based on the LAMOST DR10/DR11. This way we might be able to estimate ages in a more accurate way.

\section{Results} 

In this section, we show the distribution of the final predicted ages in sky Galactic coordinate . We also provide the distribution of the predicted age relative error analysis carefully and compare it with the literature in order to show our method and age results are reasonable. Furthermore, we compare our predicted ages with those of open clusters. Although they do not constitute absolute benchmark, we can still use them to search for a final statistical uncertainty. We also show the correlation between [C/N] and the age of RGB stars, as well as the spatial distribution of the predicted ages.

\subsection{Final stellar age distribution}

\begin{figure}
  \centering
  \includegraphics[width=0.45\textwidth]{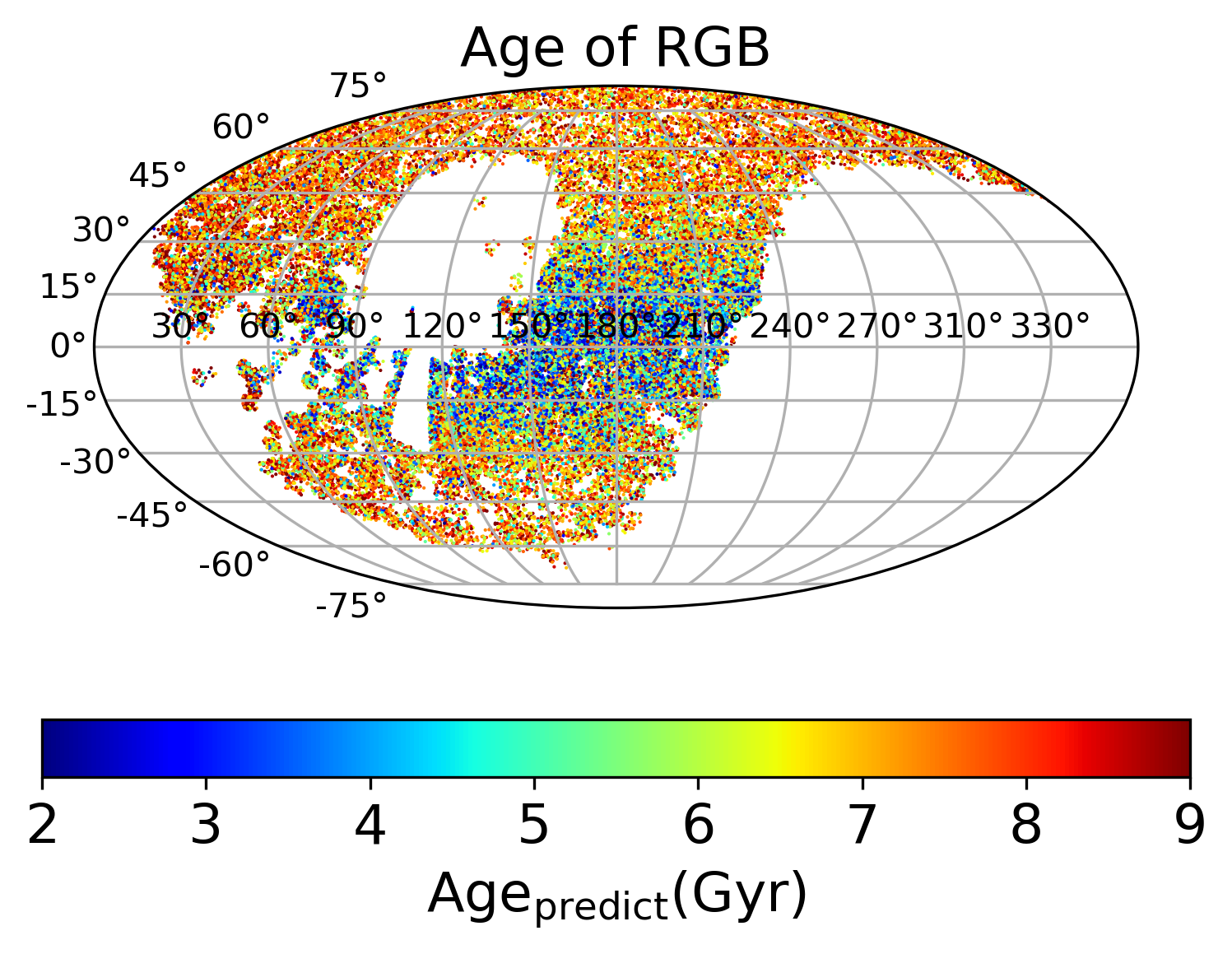}
  \caption{Distribution of the predicted RGB stars ages in Galactic coordinates.}
  \label{Distribution_Age_of_RGB}
\end{figure}

As mentioned, six basic stellar features or the input parameters ([$\alpha$/Fe], [C/Fe], T$_{eff}$, [N/Fe], [C/H], log g) are used to determine the ages of 596,116 RGB stars in the GBDT framework. Its distribution is shown in the sky Galactic coordinate in Figure~\ref{Distribution_Age_of_RGB}, with age represented by colour. One can readily see that, as expected, at lower latitudes there are mostly younger stars, while at higher latitudes the stars tend to be older. 

\subsection{More stellar age determinations comparisons}

\begin{figure*}
  \centering
  \includegraphics[width= 1 \textwidth]{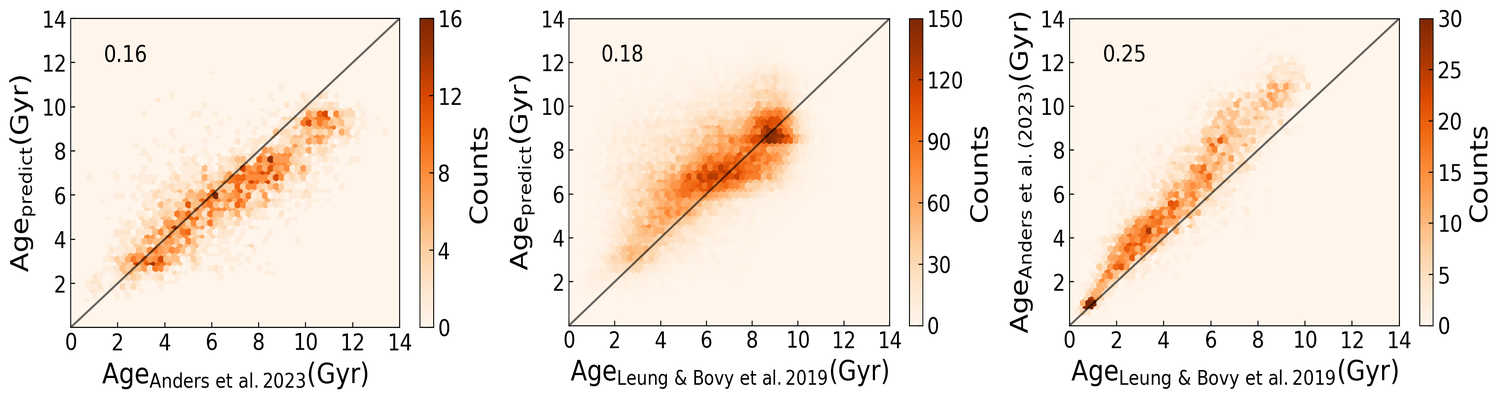}
  \caption{Comparison of the ages predicted in this paper with the results of \citet{2023arXiv230408276A} (left panel), \citet{2019MNRAS.483.3255L} (middle panel), and the results of \citet{2023arXiv230408276A} compared with those of \citet{2019MNRAS.483.3255L} (right panel). Colors indicate the number of stars in the bin. The black line is used to guide the eye for the perfect matching. The number in the upper left corner of each subplot is provided by the median relative error.}
  \label{Other_age_comparision}
\end{figure*}

In Figure~\ref{Other_age_comparision}, we compare our stellar ages predicted using the GBDT algorithm with the results of other age prediction studies. The age determinations from two previous independent works are based on APOGEE DR17 RGB giants, and employing machine learning methods and similar training samples. \citet{2023arXiv230408276A} used the eXtreme Gradient Boosting (XGBoost) algorithm, which is applied to the APOGEE-Kepler \citep{2021yCat..36450085M} observations of 3,060 RGB and red clump stars having high quality asteroseismic ages. \citet{2019MNRAS.483.3255L} presented the astroNN algorithm, suited to determine not only stellar atmospheric parameters, but also distances and ages from APOGEE spectra. The astroNN age estimation follows the procedure outlined in \citet{2019MNRAS.489..176M}, and the training set is composed of the APOKASC-2 catalog \citep{2018ApJS..239...32P} and a sample of  96 low metallicity stars from \citet{2021NatAs...5..640M}. This method depends on the relationship between the surface metallicity of RGB stars and their MS mass determined at the time of the star's First Dredge Up (FDU). 

\begin{figure*}
  \centering
  \includegraphics[width= 0.7\textwidth]{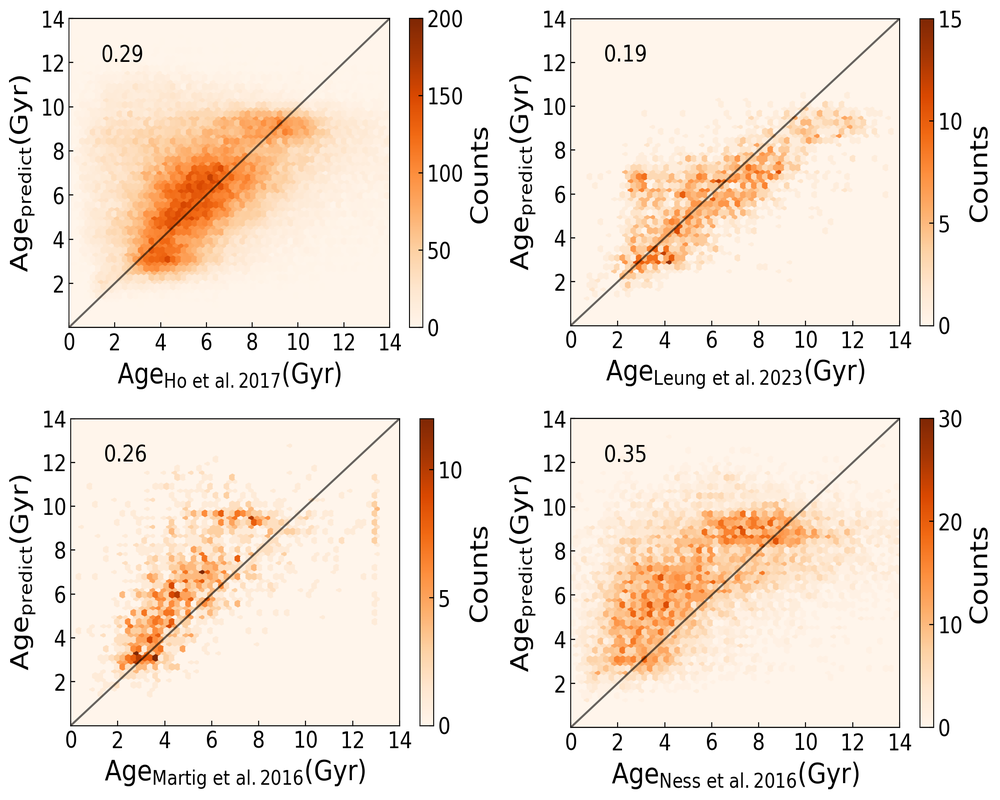}
  \caption{The ages derived from the predictions in this paper are compared with the stellar ages derived in other studies. The results of \citet{2017ApJ...841...40H, 2023MNRAS.522.4577L, 2016yCat..74563655M} and  \citet{2016ApJ...823..114N} are compared with our result from the top left panel to the bottom right panel. The relative error is shown in the left-top corner.}
  \label{Age_comparision_GBDT}
\end{figure*}

As seen in the left two panels of Figure~\ref{Other_age_comparision}, our results are more compatible with \citet{2019MNRAS.483.3255L} than with \citet{2023arXiv230408276A}. The median relative random errors are similar, amounting to $\sim$16 $\%$, but the systematics are not the same. In order to highlight the systematics we compare the ages of \citet{2019MNRAS.483.3255L} with  \citet{2023arXiv230408276A}. The results indicates that the systematics of both studies are evident, and the ages from \citet{2023arXiv230408276A} are systematically larger than the  ones from \citet{2019MNRAS.483.3255L}, in the average  of about 1$-$2 Gyr.

The age estimate from \citet{2023arXiv230408276A} is about 1.2 times smaller than the age estimate from astroNN, and our age estimate agrees relatively well with the age estimate from astroNN. Systematic biases  are unavoidable and very hard to quantify especially in the case of RGB stars and  are related to the physics assumptions about mass loss, mixing, rotation and so forth. There are also sizeable statistical uncertainties \citep{2015ASSP...39..167N}. 

We we compare our predicted ages with additional studies, as  Figure~\ref{Age_comparision_GBDT}, which, together with Figure~\ref{Other_age_comparision}, validates our approach. From the top left panel to the bottom right panel of Figure~\ref{Age_comparision_GBDT} we compare our predicted ages with those of \citet{2017ApJ...841...40H, 2023MNRAS.522.4577L, 2016yCat..74563655M} and \citet{2016ApJ...823..114N}, respectively. The figure shows that for stars having an age estimate the comparison is reasonable. The errors range from 19 $\%$ to 35 $\%$. 

\begin{figure*}
  \centering
  \includegraphics[width= 0.9\textwidth]{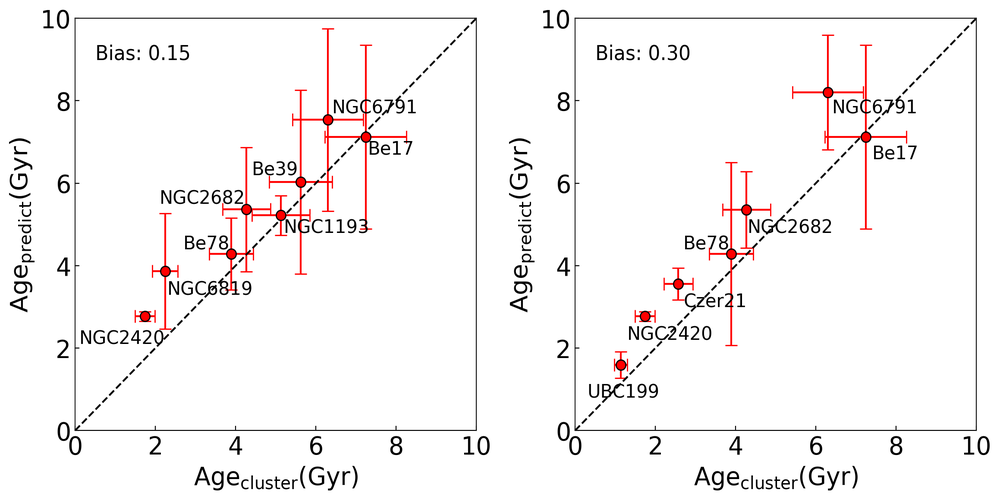}
  \caption{Basing on the catalog provided by \citet{2020A&A...640A...1C}, we compare the literature values of open clusters with our predictions. The horizontal error bars is the uncertainty quoted by \citet{2020A&A...640A...1C}, and the error bars in the y axis are provided by the standard deviation of the cluster members, which is selected by the OC clustering distribution. The final bias for open cluster is shown in the top. The left panel cluster member stars are selected by the steps of Figure~\ref{selection_cluster_NGC1193}, and the right panel member stars are obtained by directly cross-matching with the sky coordinate of the cluster member stars provided by \citet{2020A&A...640A...1C}.}
  \label{comparision_cluster}
\end{figure*}

\begin{figure*}
  \centering
  \includegraphics[width= 0.75\textwidth]{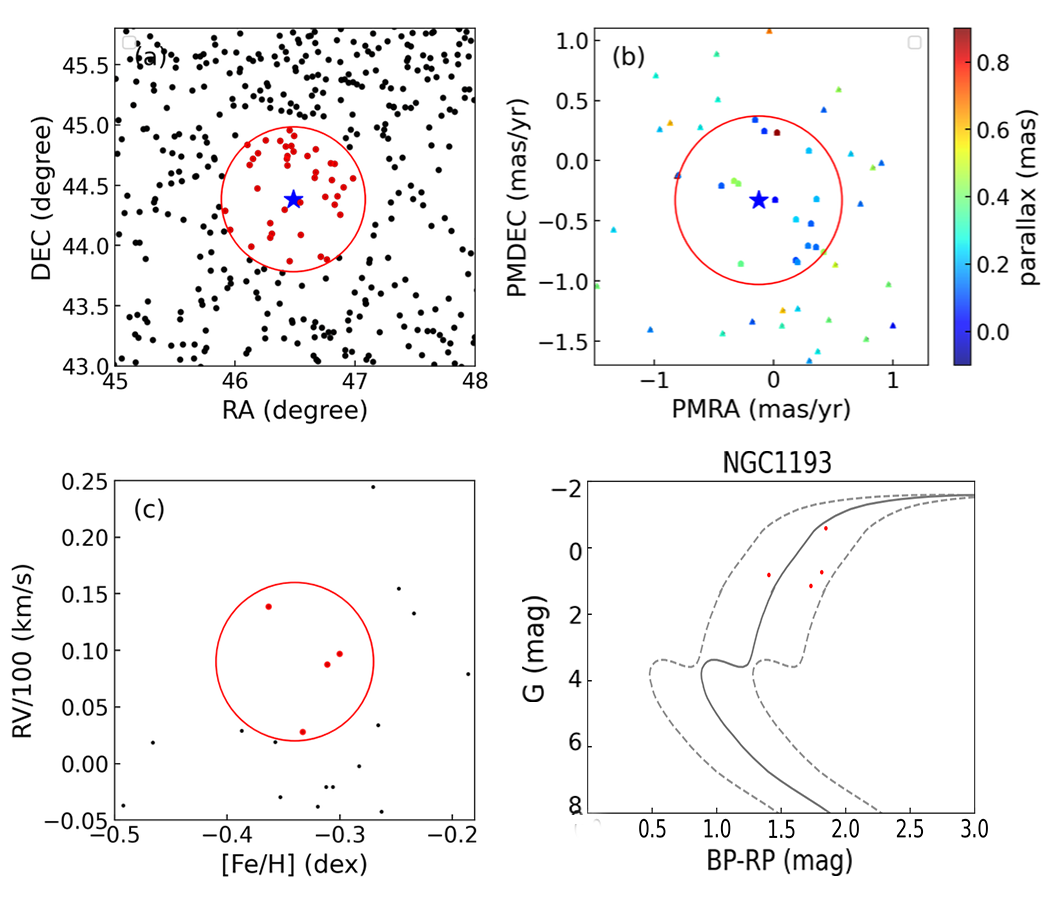}
  \caption{The selection process for membership of OC is illustrated using NGC 1193 as an example. The alphabetical order in each panel represents the order in which the cluster members were selected. The blue pentagram denotes the sky equatorial coordinate and proper motion of NGC 1193. The stars in the red circle indicate the members we have selected at this step. The solid grey line is the isochrone for NGC 1193 with 5.13 Gyr, and the two dashed lines indicate a deviation of 0.4 mag.}
  \label{selection_cluster_NGC1193}
\end{figure*}

We remind the reader that \citet{2017ApJ...836....5H} used Cannon to construct predictive models for LAMOST low resolution spectra. To convert [C/M] and [N/M] to mass and age, they used formulas for the coefficient representations in their Tables A2 and A3 of \citet{2016yCat..74563655M}. These formulas are in turn derived from asteroseismic mass measurements of stars using [C/M] and [N/M] measurements. However, we underline that these relations only apply to a certain range of labeled values, since they used the range-bound eigenvalues, which prevents them from estimating the masses and ages of low metal abundance outlier stars in the LAMOST data. 

\citet{2016ApJ...823..114N} also used the Cannon method, using masses measured directly from APOGEE spectra (R $\sim$ 22,500) and ages estimated by isochrones fitting, independent on carbon or nitrogen. Both the \citet{2016yCat..74563655M} relation and the \citet{2016ApJ...823..114N} catalog were calibrated on the same stellar sample (APOKASC). As highlighted in \citet{2016yCat..74563655M}, the mass (and age) estimates  for single stars must be used with great caution. For individual stars, the surface carbon and nitrogen abundances may not always reflect their current stellar masses, and it is possible that the presence of binary companions may have an impact. Therefore, their method may be more suited for statistical studies of large samples of stars and for comparing the properties of different populations. Recently, \citet{2023MNRAS.522.4577L} adopt a variational encoder-decoder method to create a public available stellar age catalog of 140,000 stars from APOGEE DR17, again using the age values of \citet{2021yCat..36450085M} as a training sample. When compared with the ages of astroNN, the overall systematics is about 25 $\%$. We find that the random errors from the comparison between ours and their ages amounts to  about 19 $\%$, but the systematics is not so relevant for ages below 9 Gyr. 

Furthermore, Figure~\ref{comparision_cluster} (left panel)  shows a comparison of the age estimates of our sample with the open cluster (OC) ages from \citet{2020A&A...640A...1C}. They used an artificial neural network trained with a mixture of simulated and real data (mainly from \citet{2019A&A...623A.108B}) to estimate the distance, age, and interstellar reddening of about 2,000 clusters from  Gaia DR2. The 8 open clusters were selected by position, line-of-sight velocity, and [Fe/H] (referenced from \citet{2021MNRAS.504..356D} and \citet{2020AJ....159..199D}). The horizontal error bars are derived from the uncertainty cited by \citet{2020A&A...640A...1C}, and the error bars in the y direction are provided by the standard deviation of the cluster members we selected. In the left panel, we mainly use sky coordinate, proper motion, parallax, and isochrone fitting for the selection. We would like to note also that we have not enough metallicity and radial velocity information. Finally, the number of members of the above open clusters ranges from 2 to 4 stars.

\begin{figure}
  \centering
  \includegraphics[width=0.48\textwidth]{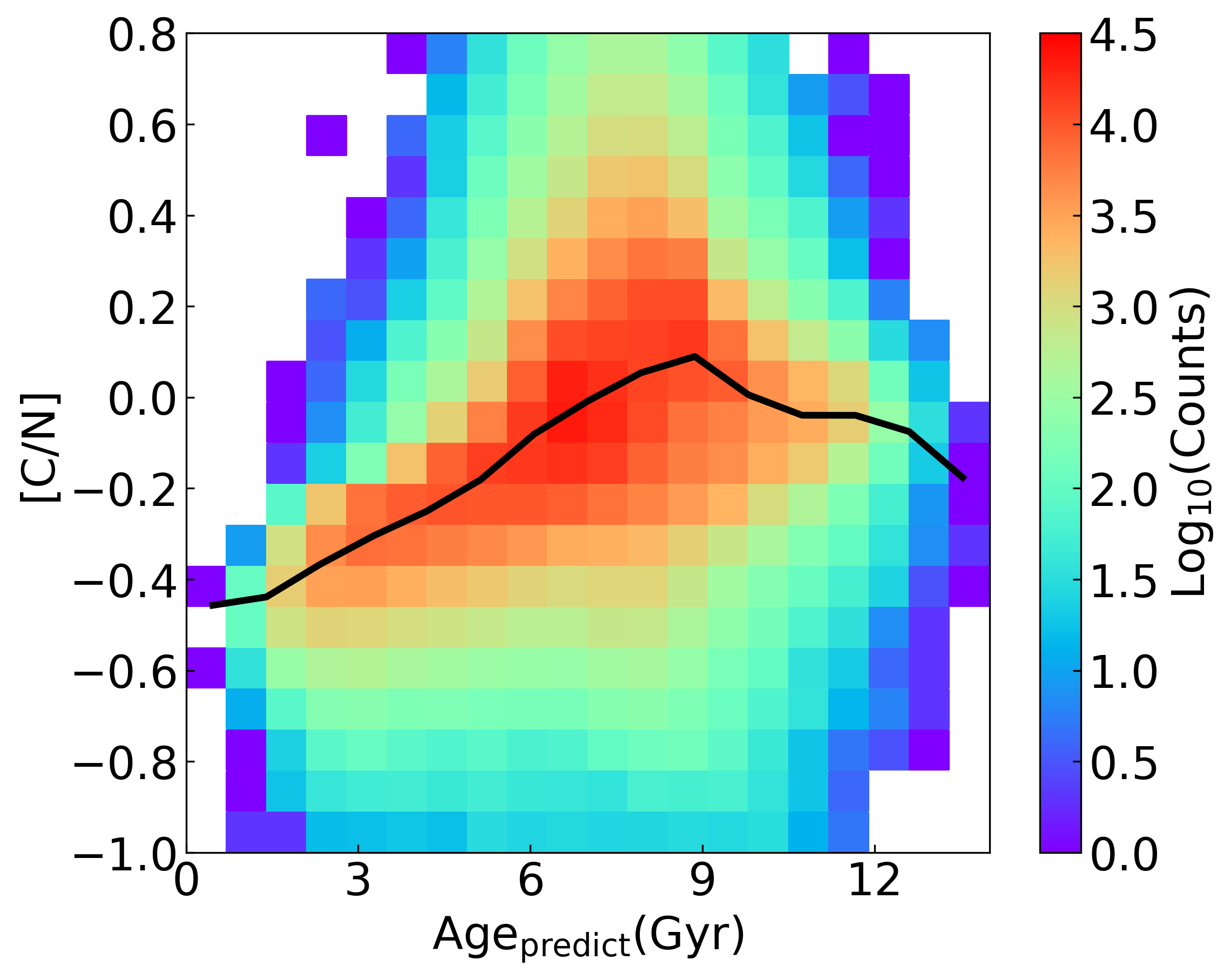}
  \caption{Relationship between the predicted age and the [C/N] ratio. The colours indicate the logarithmic values of the number of stars. The black line is the median value of the [C/N] ratio in each bin.}
  \label{Age_C_N_Counts}
\end{figure}

\begin{figure*}
  \centering
  \includegraphics[width=0.9\textwidth]{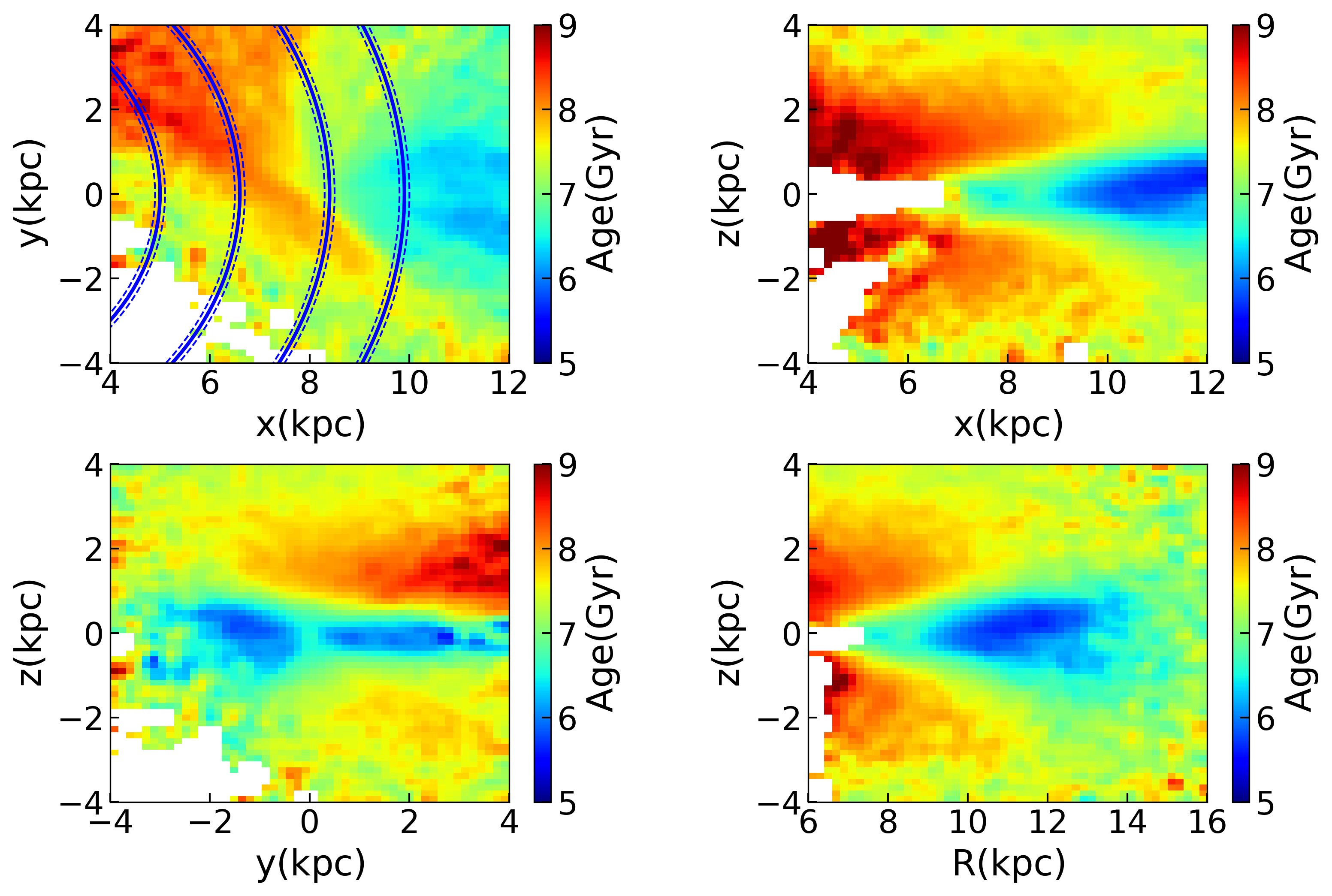}
  \caption{The spatial distribution of the 596,116 red giants for which we predict the age, here color-coded. The lower right panel uses Galactic cylindrical coordinates, while the first three panels are Cartesian coordinates. The blue lines in the top left panel indicate the four spiral arms from \citet{Reid14}.}
  \label{XYZR_age_GBDT}
\end{figure*}

One can readily see that our predicted ages matched well with the ages of the open clusters, with an overall bias is about 15 \% compared the OC values. The systematics of NGC6819 are slightly larger, around 1.63 Gyr; Berkeley 39 has the largest error bars. Except for Berkeley 39, NGC 6791 and Berkeley 17, the other open clusters have errors below 2 Gyr. We can also notice that the errors for older open clusters are overall larger than those for younger open clusters.  

Spatial position, proper motion, radial velocity, and [Fe/H] are used as selection criteria for clusters' member stars. In case some stars do not have metallicity or radial velocity then we choose other criteria to select the memberships. In general, the selected member stars have a spatial position difference from the cluster of less than 1 degree, proper motion difference within 1.2 mas/yr, and [Fe/H] variation within 0.05 dex.  The number of selected stars for each cluster are as follow: NGC 2420: 3, NGC 6819: 2, Berkeley 78: 2, NGC 2682: 3, NGC 1193: 4, Berkeley 39: 2, NGC 6791: 3, Berkeley 17: 4.

In the right panel of Figure~\ref{comparision_cluster}, we select the cluster member stars by cross-matching them directly with the equatorial coordinates of the cluster's star catalog, with a maximum error of no more than 1 arcsec. The final number of open cluster member stars ranges from 2 to 9. The overall bias compared to the age values of the open clusters is about 30$\%$.

Figure~\ref{selection_cluster_NGC1193} shows the process of selecting cluster membership in our predicted sample using NGC 1193 as an example. First, we constrained the spatial position within a range of no more than 0.7°. Second, we further limited the proper motion to no more than 1 mas/yr. Third, we restricted the [Fe/H] values within a range of no more than 0.03 dex, since member stars of a cluster are expected to originate from the same molecular cloud and hence  share similar metallicity. Finally, comparing the selected member stars of cluster to the PARSEC isochrone as a validation \citep{2012MNRAS.427..127B}.

\subsection{Stellar age-[C/N] ratio relation}

It has been suggested long ago to use the  [C/N] ratio from spectra as a proxy of RGB stellar ages : during the MS, the CNO cycle in the star core determines the final relative abundances of carbon and nitrogen elements. The first dredge-up takes place as stars in the MS phase evolve to the RGB branch stage, where the surface convective layer expands downward to the fusion material layer, and the material in the core is moved up to the surface by convective mixing. The depth of the convective envelope and the [C/N] ratio in the core are determined by the mass of the star. The mass and age of stars on giant branches are therefore closely related, which also implies that the [C/N] ratio can be used to infer stellar ages \citep{2015A&A...583A..87S,2019MNRAS.484..294D}. 

In Figure~\ref{Age_C_N_Counts}, we show the [C/N] ratio versus the predicted age. We calculate the median value in the star bins to show [C/N] as a function of the predicted age. For stars with ages less than 9 Gyr, the stellar age correlates well with the carbon to nitrogen ratio, as expected. For stellar ages larger than or equal to 9 Gyr, the ages seem to show an inverse correlation, but the number of stars in this age range is much smaller in our RGB sample, so this trend might can be spurious. Using APOGEE full-spectrum data, \citet{2019MNRAS.489..176M} applied a neural network-based model in an attempt to improve age estimation by examining the more complex relationship between age and surface abundance. This approach avoids the dependence of age estimates on the determination of individual abundance in a spectrum. They also point out that the age of sample stars may exceed the limit (10 Gyr) beyond which the spectral information no longer appears to be valid for age prediction.

\subsection{Spatial distribution of stellar ages}

In Figure~\ref{XYZR_age_GBDT} we present the spatial distribution of the 596,116 RGB age distribution in Cartesian coordinate, the redder the regions the older the median age of the stars. The left three panels clearly show the age is increasing with the height from the plane. From the right two panels, it seems that the disk flaring can be detected. As expected, younger thin disk, older thick disk, and age transitions between halo components are naturally revealed here, and we look forward to using these stars to investigate more in details the outer disk.

\section{Discussion}

\begin{figure}
  \centering
  \includegraphics[width=0.45\textwidth,height=1\textwidth]{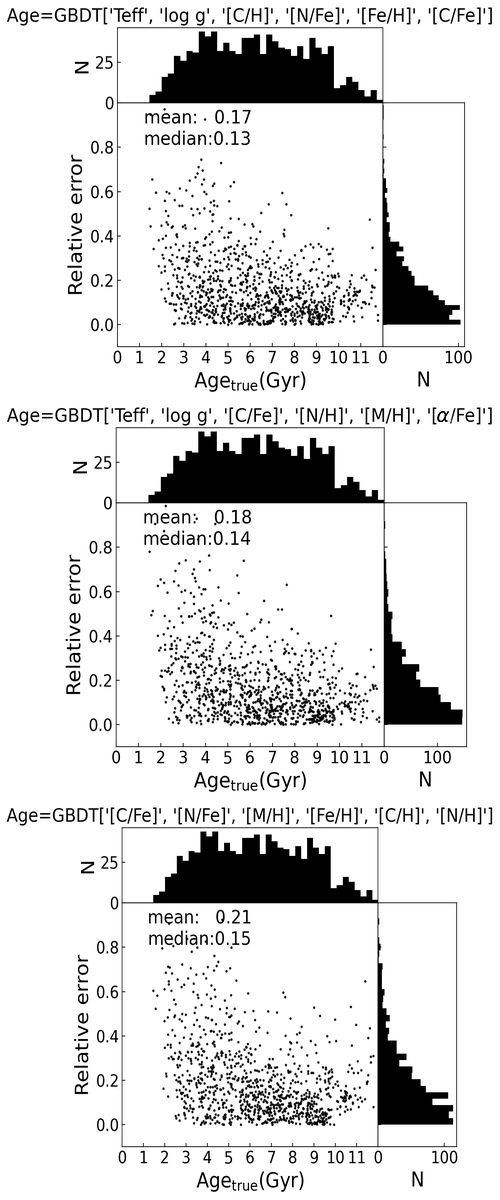}
  \caption{The stellar age predicted by other combinations of stellar parameters randomly chosen. The combinations of parameters from top to bottom are (T$_{eff}$, log g, [C/H], [N/Fe], [Fe/H],[C/Fe]), (T$_{eff}$, log g, [C/Fe], [N/H], [M/H], [$\alpha$/Fe]), ([C/Fe], [N/Fe], [M/H], [Fe/H], [C/H], [N/H]). The mean and median relative errors of the predicted ages are shown in the top left corner of three panels. It is clearly shown that the age can be determined by a variety of stellar parameters.}
  \label{Six_prediction}
\end{figure}

\subsection{Age determination of other random parameter combinations}

As a method test now we randomly choose six parameters to predict age. As shown in Figure~\ref{Six_prediction}, displaying threes examples for the training sample results with three different combinations of parameters. As can be seen from the distribution of age relative errors and the mean and median age relative errors in the Figure, the mean relative errors are 17$-$21$\%$, and the median values are all below 15 $\%$. Therefore, it is feasible to use other combinations of stellar parameters for age prediction, so that our model can achieve a good accuracy. If we used a training dataset with a larger sample size and better quality, the accuracy for our prediction results would be improved. Improving the prediction accuracy can also be done by starting with the astero-seismology parameters in the future, since it  has been shown that age predictions with astero-seismology parameters (e.g., $\nu_{max}$ and $\Delta$$\nu$) have good accuracy.

\subsection{Possible applications}
These tests in tandem with Figure~\ref{Feature_prediction} imply that many different parameters are correlated, as mentioned also in \citet{Ting2022}, which shows that many elemental abundances can be predicted from other parameters such like [Fe/H] and [Mg/Fe] (or [Fe/H] and age). 

\citet{Ting2022} also showed that cross-element correlations are more effective probes of a hidden structure than dispersion. In fact, some critical information of the elements can not be simply inferred from the metallicity and $\alpha$ elements and the residual correlation structure of the abundance trend might be important for the stellar evolution studies, interstellar mixing and the disk merger history.  

The metallicity and chemical abundance are produced by the core-collapse supernovae (e.g., $\alpha$/Fe), Type Ia supernovae ([Fe/H]), asymptotic giant branch (AGB) stars (Ce), neutron star mergers (Eu/Fe), etc.), one could find more details in \citet{Weinberg2019, Weinberg2021, Ting2012, Ting2017b}. Age is an extremely valuable tracer for the stellar evolution history and dynamical evolution of the Milky Way disk. As a prospect, we suggest one could use this sample to statistically investigate about the disk structure such like flaring and warp using mono-age populations and disk heating history based on action angles (Wang H.-F., et al. in prep; and also \citep{Frankel2020}).

\section{Conclusions}

In this study, we carefully selected a sample of RGB stars in APOKASC-2 as  training sample and  then train the age prediction model using GBDT machine learning method aiming at estimating the ages of 596,116 RGB stars of LAMOST DR8. The correlation between stellar eigen-parameters and stellar ages is analyzed using the GBDT algorithm. We select the top six feature parameters in the correlation ranking with age to train the age prediction model, and the median absolute model error can reach 0.72 Gyr and the median relative error can reach 11.6$\%$.

After comparing the predicted ages with the results of other works such as Gaia and APOGEE, we find that the overall trend is good and the median relative error of the stellar age comparison is around 16 $\%$. When comparing with the open clusters, our predicted ages are in good agreement with their ages, the final uncertainty being about 15$-$30 $\%$.

As a validation of the method, the linear relationship between the reliably predicted age and [C/N] and the expected vertical age gradient of the Galactic disk are also reproduced. Moreover, we find that our method is not only suitable for age determination, but also can be applicable for other atmospheric parameters. 

We also discuss the effects of different combinations of input feature parameters for age prediction. The mean relative errors are from 17$-$21$\%$, and the median relative errors are all below 15$\%$. It is worth exploring more for the stellar parameters including basic stellar parameters, abundance parameters, age and even distance and extinction, in the future.

\section*{Acknowledgements}

We would like to thank the anonymous referee for his/her very helpful and insightful comments. HFW is supported in this work by the Department of Physics and Astronomy of Padova University though the 2022 ARPE grant: {\it Rediscovering our Galaxy with machines.} The Guo Shou Jing Telescope (the Large Sky Area Multi-Object Fiber Spectroscopic Telescope, LAMOST) is a National Major Scientific Project built by the Chinese Academy of Sciences. Funding for the project has been provided by the National Development and Reform Commission. LAMOST is operated and managed by National Astronomical Observatories, Chinese Academy of Sciences. This work has also made use of data from the European Space Agency (ESA) mission {\it Gaia} (\url{https://www.cosmos.esa.int/gaia}), processed by the {\it Gaia} Data Processing and Analysis Consortium (DPAC, \url{https://www.cosmos.esa.int/web/gaia/dpac/consortium}). Funding for the DPAC has been provided by national institutions, in particular the institutions participating in the {\it Gaia} Multilateral Agreement.

\appendix

\section*{Predicted other stellar chemical parameters}

\begin{figure*}[!h]
  \centering
  \includegraphics[width=0.9\textwidth]{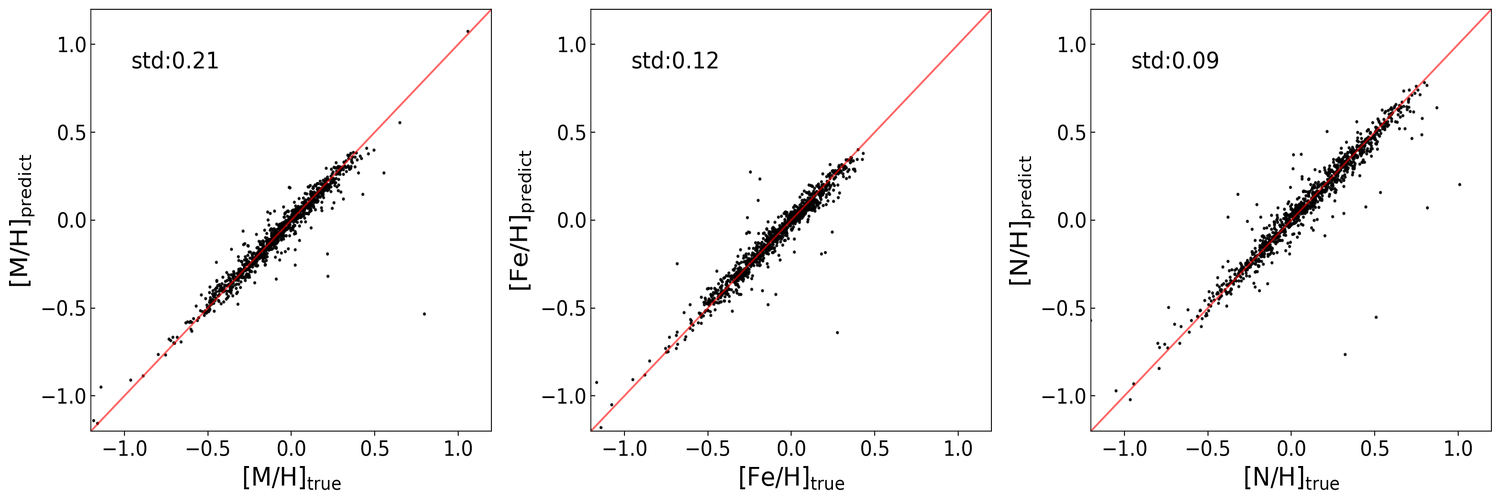}
  \caption{Predictions for the remaining {\bf 3 stellar features ( [M/H], [Fe/H], [N/H])} on the basis of the 6 stellar features of the predicted age. The dispersion is shown in the left-top corner of each panel.}
  \label{Feature_prediction}
\end{figure*}

To illustrate the correlation between different feature parameters, we use the method and 6 stellar parameters described above to predict the remaining 3 feature parameters, and compare the predicted values with the original parameters (see Figure~\ref{Feature_prediction}). In all panels of Figure~\ref{Feature_prediction}, the scatter points representing the stellar parameter values are almost uniformly distributed on both sides of the red line, indicating that our model is able to predict the stellar feature parameters robustly. The well-predicted results also show again that there is indeed a good consistency between these stellar feature parameters. The dispersion in Figure~\ref{Feature_prediction} (left corner values) also evidence that these predicted parameters are reasonable considering the uncertainties of the stellar parameters for any survey data-sets. 

\begin{figure*}[!h]
  \centering
  \includegraphics[width=0.9\textwidth]{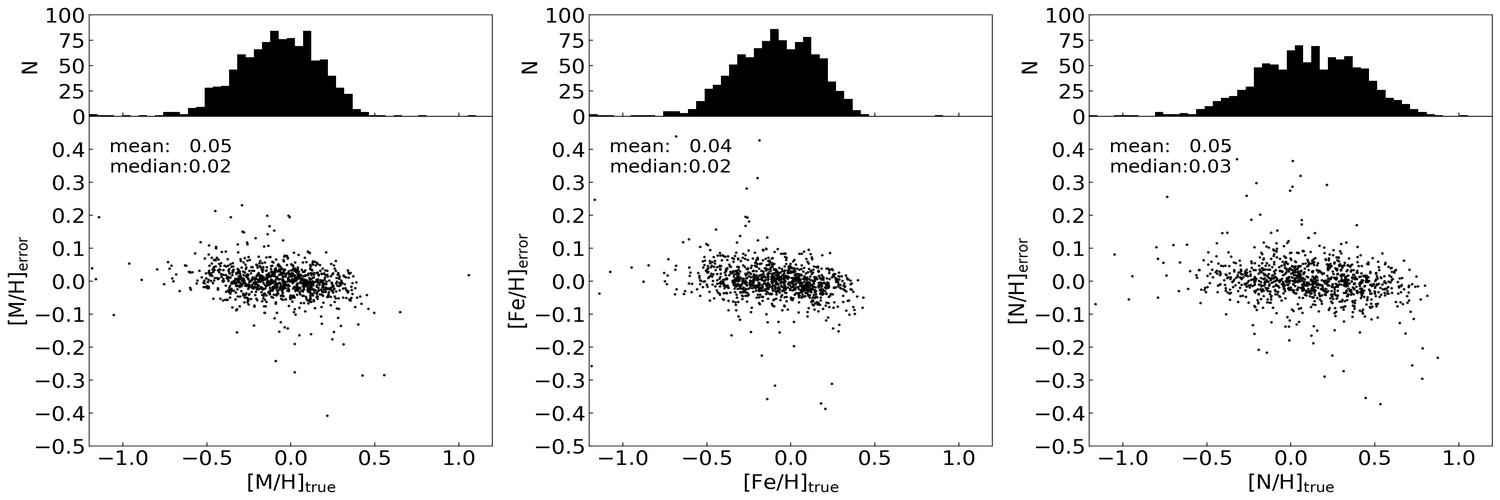}
  \caption{The error distribution of the predicted features is arranged in the same order as in Figure~\ref{Feature_prediction}. The histogram indicates the distribution of the number of stars over the values of the characteristic parameters. The predicted feature parameters obtained are in agreement with the true features for the overall trend with small relative errors.}
  \label{Feature_prediction_error_GBDT}
\end{figure*}

In Figure~\ref{Feature_prediction_error_GBDT} we show the error distributions for the 3 characteristic parameters as complement to the Figure~\ref{Feature_prediction}, implying that our model has a certain degree of reliability in predicting these correlated feature parameters. Although there is some dispersion, we expect that the method may exhibit improved performance in parameter prediction when the training sample becomes sufficiently large and precise enough in the future.
In the meantime, we believe these parameters precision might be improved by finding the feature parameters that are more correlated with them in the model training, so that our method would be surely working not only suitable atmospheric parameters but also stellar age. 

\end{document}